\documentclass[twocolumn,showpacs,preprintnumbers,amsmath,amssymb]{revtex4}

\usepackage[T1]{fontenc}
\usepackage{lmodern}
\usepackage[utf8]{inputenc}
\usepackage[english]{babel}
\usepackage{graphicx}
\usepackage{dcolumn}
\usepackage{bm}
\usepackage{color}

\begin{document}

\title{Molecular dynamics simulation of growth of Cu nanoclusters from Cu-ions in a plasma}

\author{Alexey A. Tal}
\email{aleta@ifm.liu.se}
\affiliation{Theory and Modeling, IFM-Material Physics, Linköping University, SE-581 83, Linköping, Sweden}
\affiliation{Materials Modeling and Development Laboratory, National University of Science and Technology 'MISIS', 119049, Moscow, Russia}  
\author{E. Peter Müger}%
\affiliation{Theory and Modeling, IFM-Material Physics, Linköping University, SE-581 83, Linköping, Sweden}%
\author{Igor A. Abrikosov}
\affiliation{Theory and Modeling, IFM-Material Physics, Linköping University, SE-581 83, Linköping, Sweden}%
\author{Nils Brenning}
\affiliation{Plasma \& Coatings Physics Division, IFM-Material Physics, Linköping University, SE-581 83, Linköping, Sweden}%
\affiliation{Division of Space \& Plasma, School of Electrical Engineering, Royal Institute of Technology, SE-10 044, Stockholm, Sweden}%
\author{Iris Pilch}
\affiliation{Plasma \& Coatings Physics Division, IFM-Material Physics, Linköping University, SE-581 83, Linköping, Sweden}%
\author{Ulf Helmersson}
\affiliation{Plasma \& Coatings Physics Division, IFM-Material Physics, Linköping University, SE-581 83, Linköping, Sweden}%

\begin{abstract}

A recently developed method of nanoclusters growth in a pulsed plasma is studied by means of molecular dynamics. A model that allows one to consider high-energy charged particles in classical molecular dynamics is suggested, and applied for studies of single impact events in  nanoclusters growth. In particular, we provide a comparative analysis of the well-studied inert gas aggregation method and the growth from ions in a plasma. The importance to consider of the angular distribution of incoming ions in the simulations of the nanocluster growth is underlined. A detailed study of the energy transfer from the incoming ions to a nanocluster, as well as the diffusion of incoming ions on the cluster surface is carried out. Our results are important for understanding and control of the nanocluster growth process.
\end{abstract}

\maketitle

\section{\label{sec:level1}Introduction}

Metal nanoclusters (NC) have been in focus in many experimental and theoretical investigations because of their size dependent mechanical, electrical and optical properties \cite{baletto05}. Owing to them, NCs are widely used in many applications like catalysis \cite{henry98, cuenya10}, biomedical or photovoltaic \cite{garcia11}.  Control of NC properties such as size and morphology is important for all applications. A deep understanding of the growth process has already been achieved for some of the different techniques for NC synthesis. For example, Inert Gas Aggregation (IGA) method has been intensively studied in the literature. Baletto et al. \cite{baletto00} has shown that in IGA particle morphologies are determined by a competition between kinetics and thermodynamics. Moreover, they have shown that this competition may result in a morphology transition. However, the need to increasing productivity of NC synthesis led to development of alternative techniques for their synthesis. In particular, a novel approach, employing pulsed highly-ionized plasma has been suggested in \cite{plich13a}. The main benefit of this method is a high growth rate of 470~nm/s, which was achieved by fostering growth by collection of ions rather than growth by collection of atoms \cite{plich13b}. On the other hand, there is little fundamental understanding of the process of charged NC growth from ions.

In this method the source material for the growth of NCs is provided by sputtering a hollow cathode using high power pulses. The growth of NCs can be controlled by external parameters like the pulse parameters, the gas pressure and the gas flow. The influence of the pulsing parameters on the growth of NC was studied and a more detailed description can be found elsewhere \cite{plich13a, plich13b}. Roughly speaking, the growth process can be divided into two stages. The growth starts with the formation of dimers by three-body collisions. Small clusters are charged by interactions with ions and electrons in the plasma environment. The sign and the value of the charge is determined  by fluxes of electrons and ions. The flux of electrons is larger than the positive ion flux and clusters are thus  negatively charged. Coulomb repulsion of these charges prevents the clusters form coalescing. Due to that fact, the clusters can grow further only by collection of neutrals and positively charged ions, where the ions are accelerated towards the NC by the Coulomb interaction.

Dimer formation in a plasma environment is problematic for simulating by means of classical MD. We will thus investigate NC at sizes around 150 atoms and larger, where NCs grow by attachment of individual atoms. There are different approaches to simulate  NC growth. One way is by adding material atoms to the simulation box together with an inert gas used to perform temperature control of the NCs \cite{grochola07_2}. This type of simulations mimics the IGA growth, but classical potentials are developed for bulk materials and may be irrelevant at the nucleation stage. Another way was used by Balleto et al. \cite{baletto00} where growth from a seed was simulated. The shape of the seed was chosen to be thermodynamically favorable and then atoms were generated around the seed with thermal velocities directed towards the cluster. This type of simulation mimics only one mechanism of the growth - atom by atom growth which is relevant for IGA process, since energies of incoming atoms correspond to thermal energies. 

However, there is no commonly accepted model for the simulation of NC growth from ions in a plasma environment. In a plasma environment not only energies of the ionized fraction of the incoming particles are much higher but also the type of  collisions is different. Thus using the idea suggested by Balleto et al. we develop a model for the simulations of NC growth in a plasma and compare it to simulations of the IGA method in order to show how the plasma environment affects the growth process as compared to IGA. In the case of charged particles, kinetics will be crucially affected by the Coulomb interaction. A significant fraction of particles grown by the plasma method demonstrate a morphology unfavorable at the grown sizes. A description of this fact requires a detailed investigation of the growth kinetics.

The aim of the present work is to show how conditions in a plasma will affect the growth process as compered to the well-studied IGA method. The paper is organized as follows. In Sec. II we derive the angular distribution of incoming ions  onto the NC surface and  in Sec. III we give a detailed description of the simulations. Section IV is devoted to a discussion of obtained results and Sec. V is our conclusions.

\section{Model}
Classical approximation does not allow to account for all processes that are present in a plasma, thus we are forced to neglect the effect of thermal radiation, luminescence, thermionic emission, electron detachment and field emission of electrons. The last one is expected to be the largest contribution but we believe that these processes do not make a significant difference for the dynamics.

Thus we will consider the following model as schematically depicted in Fig.~1. A cluster has a negative charge and ions around are positively charged. The Coulomb interaction attracts the ions towards the cluster. But since ions have random thermal velocities when they start to get attracted towards the NC they fall on the cluster with different angles of incidence depending on the direction and magnitude of their initial velocity. For small energies of the incoming particles they are likely to be trapped by the NC. On the contrary, for higher energies, especially with tangential arrival, the particles may be scattered and not trapped on the NC. 

The growth speed from sizes 10 to 40 nm is shown in \cite{plich13a} to correspond to a typical electron temperature of 1.7 eV, provided that orbit motion limited (OML) theory \citep{allen92} for charging is assumed. This would correspond to a particle potential of about -4 V. However, at our gas pressure ($\sim$ 100~Pa), collision enhanced collection (CEC) of ions \citep{gatti08} will reduce this value. Furthermore, for NCs with a radius about of 1 nm as we study here, electron field emission is likely to uncharge the NC until only one electron remains.  We estimate the likely NC potential to be in the range -1 to  -4~V, and use 1 eV for incoming ions here for a case study. Thermal energy of ions is estimated to be 0.03 eV, in thermal balance with the process gas as shown in \cite{hasan13}. 

\begin{figure}[!h]
\begin{center}
\includegraphics[width=0.5\textwidth]{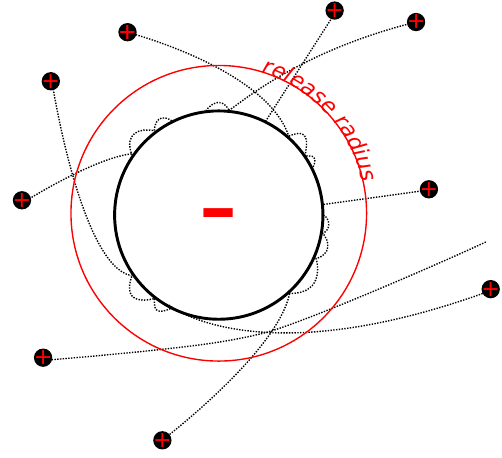}
\caption{Model of a cluster in plasma environment. Velocities and angular distribution of ions moving in Coulomb potential are calculated analytically up to the release radius. After the release boundary, the calculation of the ion path is made by molecular dynamics simulations.}
\end{center}
\end{figure}
Simulations of charged particles in classical MD is a very complex problem as charge transfer can not be taken into account explicitly. We suggest to circumvent this problem by calculating the trajectories and velocities of moving ions analytically, whereupon ions could be released near the surface of the cluster with known velocity and zero charge.

In order to find trajectories and the angular distribution of ions moving in the field of the cluster, it is needed to solve the problem of two-body interaction presented schematically in Fig.~2. The ion velocity far from the cluster is isotropic which is equivalent to a non uniform angular probability distribution for the angle $\alpha_0$ between the radius to the ion and the ion velocity:

\begin{equation}
\int \! 2\pi A \sin\alpha_0 \, \mathrm{d}\alpha_0 = 1  
\end{equation}
where A is a normalization constant. From the conservation of angular momentum it follows that:
\begin{equation}
[\vec{r_0} \times m\vec{v_0}] = [\vec{R} \times m\vec{v}] 
\end{equation}
where $\vec{r_0}$ is the distance from the center of the cluster to the initial position of the ion and $\vec{v_0}$ its corresponding velocity, $\vec{R}$ and $\vec{v}$ are the distance and velocity of the ion close to the cluster surface and $m$ is the ion mass. Let us denote by $\alpha$ an angle between $\vec{R}$ and $\vec{v}$. Then Eq. (2) can be re-written as: 
\begin{equation}
r_0v_0\sin{\alpha_0} = Rv\sin{\alpha} 
\end{equation}
Solving Eq. (3) for $\alpha_0$ we obtain:
\begin{equation}
\alpha_0 = \arcsin{\left(\frac{Rv}{r_0v_0}\sin{\alpha}\right)} 
\end{equation}
The Jacobian of the coordinate transformation is:
\begin{equation}
J=\frac{\mathrm d \alpha_0}{\mathrm d \alpha} = \frac{\frac{Rv}{r_0v_0}\cos{\alpha}}{\sqrt{1-(\frac{Rv}{r_0v_0})^2\sin^2{\alpha}}}
\end{equation}
Consequently, the angular distribution near the cluster surface is:
\begin{equation}
\int \! \frac{2\pi B 1/2(\frac{Rv}{r_0v_0})^2\sin{2\alpha}}{\sqrt{1-(\frac{Rv}{r_0v_0})^2\sin^2{\alpha}}} \mathrm d \alpha = 1
\end{equation}
where B is a normalization constant.

\begin{figure}[!]
\begin{center}
\includegraphics[width=0.4\textwidth]{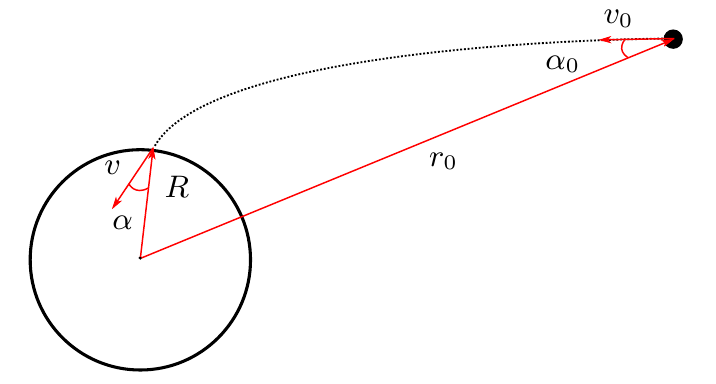}
\caption{Scheme of interaction between an incoming ion and a cluster with opposite charges. This model is used to find trajectories for incoming particles and their angular distribution. Here $\vec{r_0}$ is the distance from the center of the cluster to the initial position of the ion and $\vec{v_0}$ its corresponding velocity and an angle between them is $\alpha_0$. $\vec{R}$ and $\vec{v}$ are the distance and velocity of the ion close to the cluster surface and an angle between them is denoted by $\alpha$.}
\end{center}
\end{figure}
In Fig. 3 we show the angular distribution of incoming particles calculated from the expression (6) for four different energies of the Coulomb interaction potential: 0.03~eV (corresponding to IGA), 1~eV, 5~eV and 10~eV. The distributions for low energies like IGA or 1 eV look very similar, however, for higher energies like 5~eV and 10~eV the distributions are very different. For the case of high-energy particles (10~eV) the distribution is narrow and most of the ions bombard the cluster with small incidence angles. This is due to the fact that trajectories of low-energy particles bend around the cluster, but if the particles are fast enough they either collide with almost normal incidence or just miss the cluster. Besides, the distribution of 1 eV ions has its maximum at $\approx 45^\circ$ which means that very few particles will collide with normal or grazing incidence and most of them will come with angles around $\approx 45^\circ$. Note that even though the distribution for particles with lowest energies, 0.03~eV looks similar to the one with 1~eV, most of those particles will still have normal incidence upon the collisions with the cluster surface. Indeed, they move so slow that the interactions with the NP inside the cutoff radius are sufficient to change angles $\alpha$ to 0. Thus, conventional assumptions made in studies of NC growth in IGA should not be affected by our results. The distribution for the middle-energy particles (5~eV) has a different shape than low-energy particles but still has a maximum at $\approx 45^\circ$.

It is important to note that in our model for the angular distribution of incoming ions, only the Coulomb interaction between the cluster and ion is considered. However, when the ion gets close to the cluster surface, within the cutoff distance for the classical MD potential, it is instead treated with MD as an atom. It then gets accelerated towards the cluster and that makes the angular distribution narrower as well. This effect is larger the lower the ion’s velocity is.\newline
\begin{figure}[ht]
\begin{center}
\includegraphics[width=0.5\textwidth]{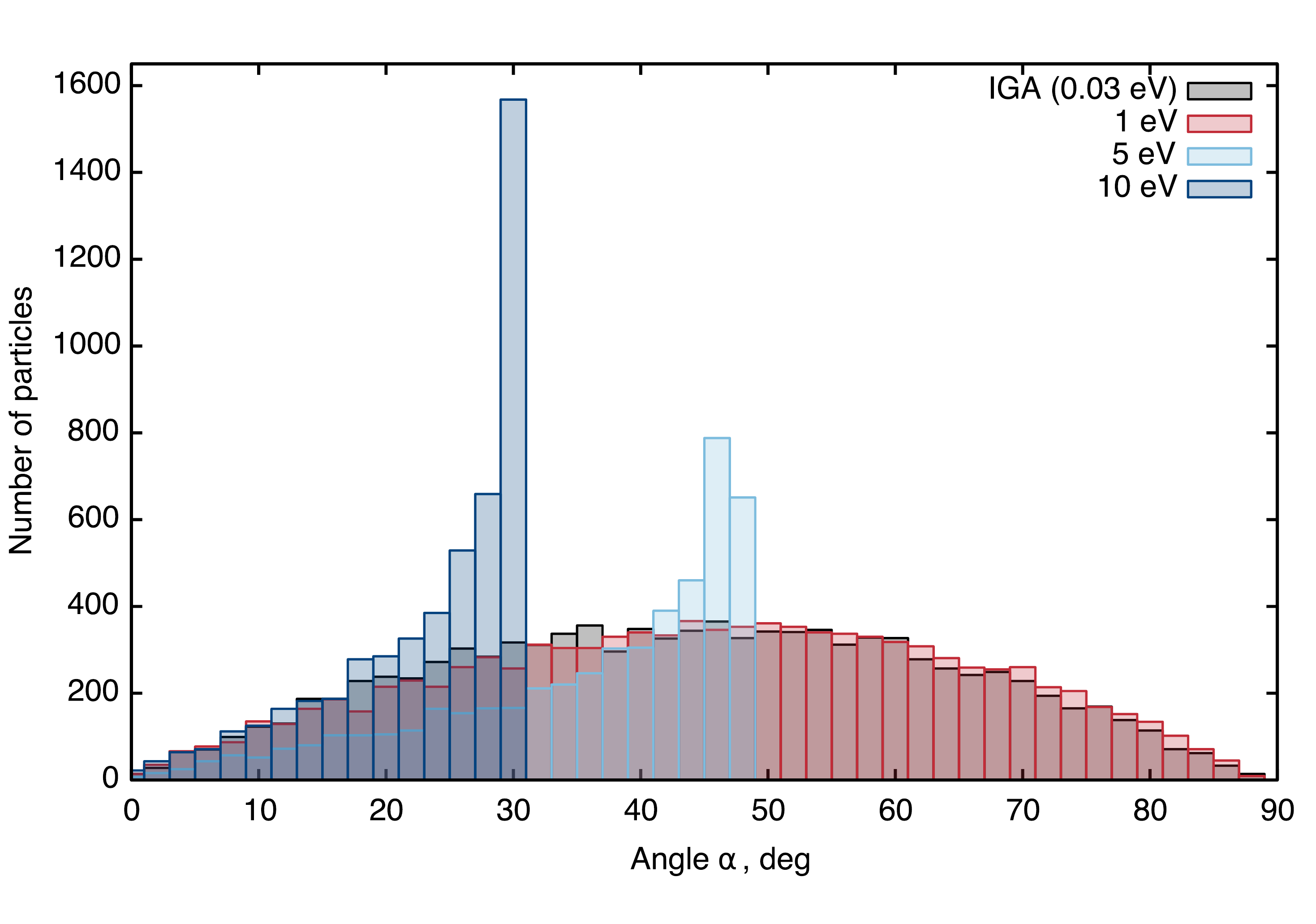}

\caption{Angular distribution for incoming particles obtained from the expression (6) for four energies IGA (0.03~eV), 1~eV, 5~eV and 10~eV. The distributions are calculated with parameters: $R = 10~nm$, $r_0 = 100~nm$, $v_0 = 300~m/s $, $m = 10^{-25}~kg $   }
\end{center}
\end{figure}

\section{Simulations details}

In this work the growth process was simulated with the embedded-atom method (EAM) potential with Foiles  parameterization \cite{foiles86}. Based on the calculation made by Baletto et al. \citep{baletto02}, we used a cluster consisting of 147 atoms with icosahedral morphology as an initial growth seed. In every simulation an incoming particle was randomly generated on a sphere with radius 13~Å. This radius was chosen to be slightly larger than the cutoff radius of the potential. Particles were generated with velocities corresponding to 1~eV directed with angle $\alpha$ (see Fig.~2). The simulations were performed using MD code LAMMPS \cite{lammps} with the velocity Verlet algorithm and a time step of 1~fs for integration in time. Previous work has shown that an inert gas environment have little or no significant kinetic effect on the growth process in IGA \cite{grochola07_1}, thus simulations were performed in NVT-ensemble and the temperature was controlled by Nosé-Hoover thermostat \cite{nose84}. The temperature of a growing NC is determined by a balance between cooling and heating processes. It can be influenced during experimental synthesis of NCs by several means: through varying the electron temperature, the plasma density, and the process gas species and pressure. For a case study we here assume 300 K. The main difference in simulations of single events in a IGA process and  a growth from ions in a plasma is that in the later case ions have higher energy and a different angular distribution.  

\section{Results}

We performed series of simulations of single atom collisions. Every simulation was initiated with one incoming atom with a velocity corresponding to 1~eV and angle of incidence $\alpha$. Fig.~4 shows snapshots from  a typical simulation where the atom colors represent their kinetic energy from the low energy shown by blue color to the high energy denoted with red color. Initially the atom has a pink color, indicating that it has rather high energy compared to the thermal energy of atoms in the cluster. Then after approaching the cutoff radius of the potential it gets accelerated and its kinetic energy increases, as illustrated with red color. The atom loses its energy in interactions with the cluster by transferring kinetic energy into heat. The heat transfer is known to affect the growth process.\newline

\begin{figure}[h!]
\begin{center}
\includegraphics[width=0.5\textwidth]{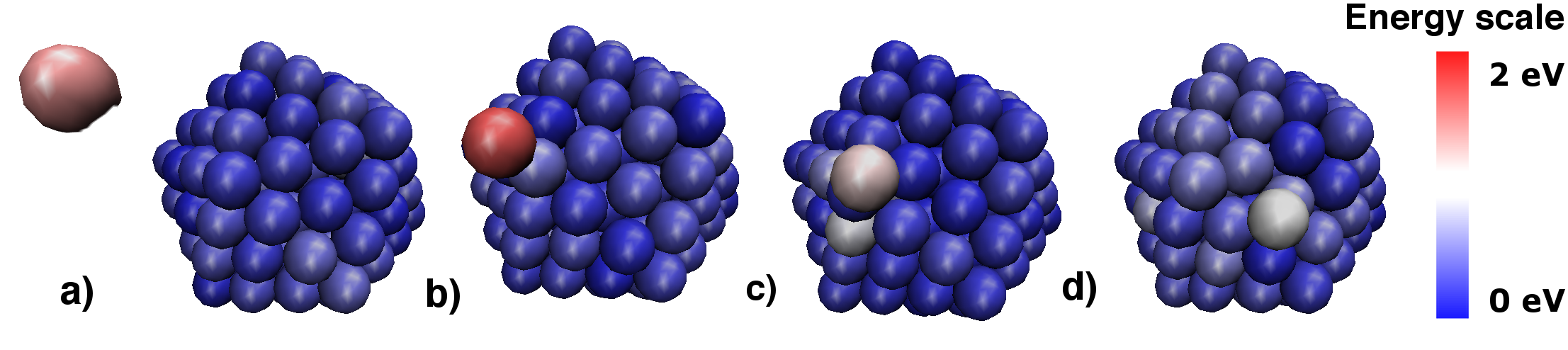}
\end{center}
\caption{Snapshots from a typical single event simulation.The color of an atom represents its kinetic energy. a) The incoming atom beyond the cutoff radius; b) the atom is inside the cutoff radius of the potential and accelerated towards the cluster. A change of its color indicates a higher energy; c) the atom transfers its energy to the cluster in collisions with the cluster atoms; d) the atom is thermalized on the surface of the cluster.}
\end{figure}

In order to understand the difference in kinetic processes during IGA and the growth in a plasma from ions a detailed characterization of heat transfer is required. In the IGA growth process incoming atoms always have normal incidence due to low kinetic energy, thus when an incoming atom approaches the cutoff radius it gets accelerated towards the cluster and collisions are always with normal incidence angle.

Fig. 5 shows how the kinetic energy changes with time for incoming particles with different incidence angles. Curves are the result of averaging over 160 independent simulations. For incoming atoms with normal incidence heat transfer is very fast and consequently leads to strong heating of a local spot. The atom with grazing incidence loses its energy in serial interactions with many atoms on the surface of the cluster which leads to broadening of the peak. In that case local heating is less significant. In the IGA process, a particle has lower velocity and consequently the energy transfer is slower since the particle approaches the cluster more slowly, that explains why the width of the peak in this case is larger than in case of high-energy particles with normal incidence.

\begin{figure}[h!]
\begin{center}
\includegraphics[width=0.5\textwidth]{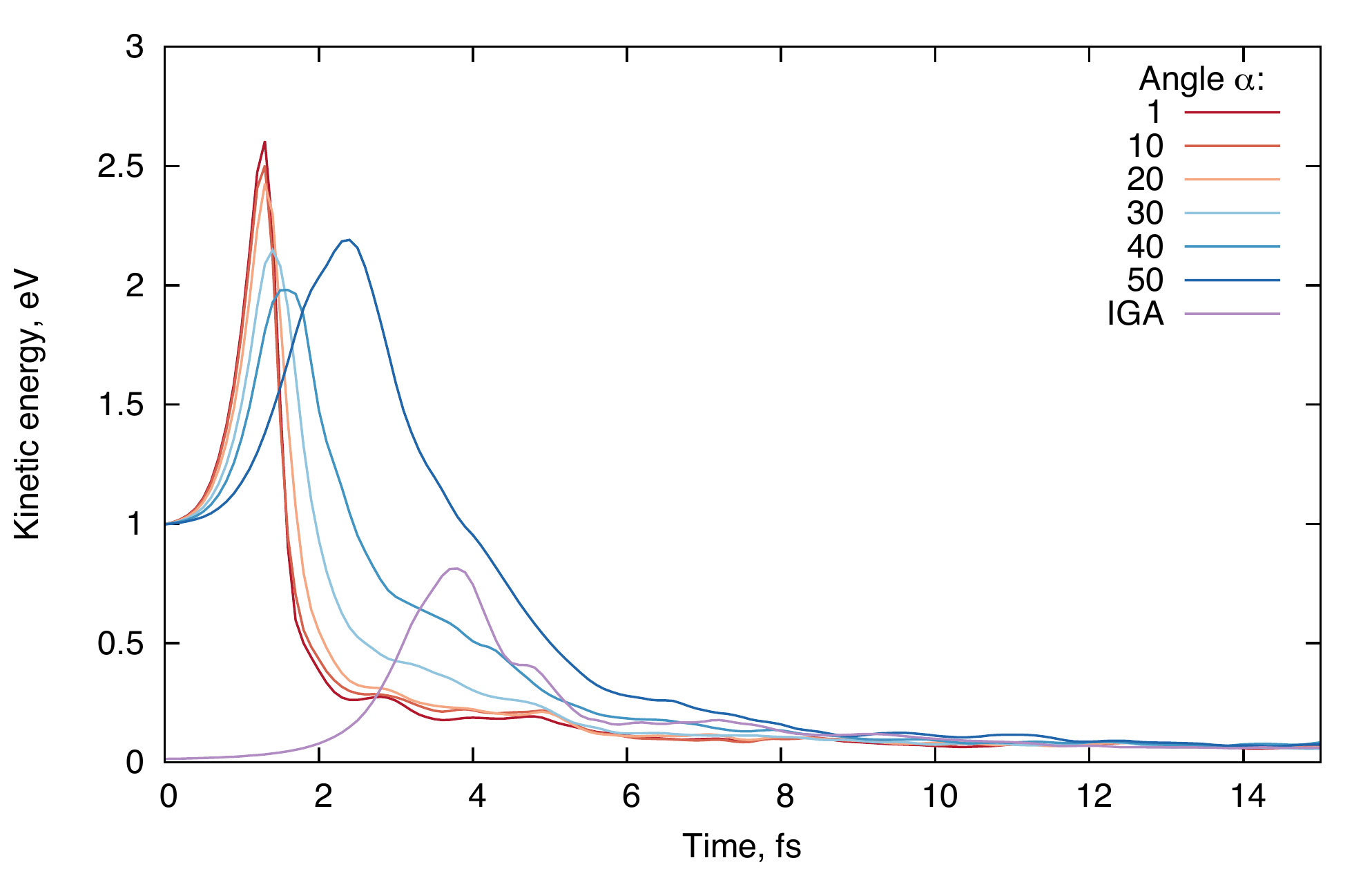}
\end{center}
\caption{Kinetic energy of incoming particles with incidence angle $\alpha$ as a function of time. Six curves correspond to high energy, 1~eV, particles for six different angles. The IGA curve corresponds to particles with thermal-energy, 0.03 eV}
\end{figure}

The difference in kinetic processes between IGA and plasma growth processes becomes clearer if one compares rates of thermalization of incoming particles. Fig.~6 shows how fast incoming atoms with different angles are thermalized. We consider a high-energy particle as thermalized when its energy has decreased 3 times. Simulations have shown that in the IGA process the angle of incoming particles does not affect the cooling rate at all. Whereas in plasma growth the dependence is nonlinear. In 
Fig.~6 this nonlinear dependence is fitted with a parabola. Each point in this figure is the result of averaging over 160 independent simulations. \newline

\begin{figure}[h!]
\begin{center}
\includegraphics[width=0.5\textwidth]{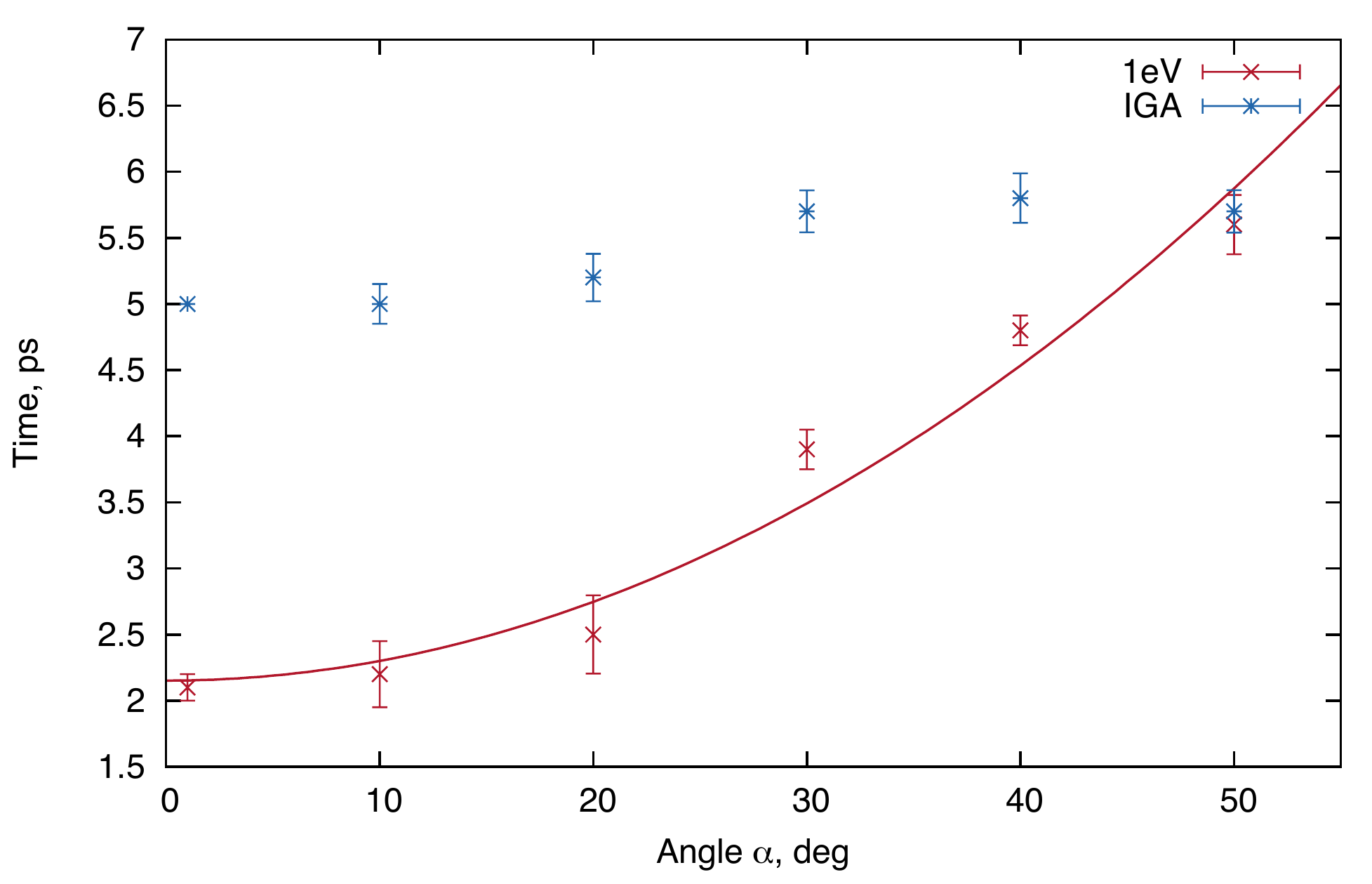}
\end{center}
\caption{Rate of thermalization as a function of incidence angle for particles in the plasma growth with energy 1~eV and particles in IGA process with thermal energy, 0.03~eV. The rate of thermalization for high-energy particles  depends nonlinearly on the incidence angle and is fitted with a parabola while the thermalization rate for the IGA process does not show a dependence on the incidence angle}
\end{figure}

The incidence angles of incoming particles considerably affect the surface diffusion. When a particle approaches the cutoff radius of the potential, it starts to interact with the cluster. This position was considered as an initial point for diffusion length calculations. Fig.~7 demonstrates how the distance from the initial position changes with time for the incoming particles. Due to interactions with the cluster the incoming particle loses all its initial energy and finally becomes thermalized, that corresponds to the plateau on Fig. 7. In the IGA process collisions are always normal, thus surface diffusion is low. On the contrary high energy particles with large incidence angles can pass up to 15~Å over the surface until they lose their initial kinetic energy. It is worth mentioning that from the viewpoint of diffusion there is no significant difference between high and low energy particles with 
normal incidence. \newline

\begin{figure}[h!]
\begin{center}
\includegraphics[width=0.5\textwidth]{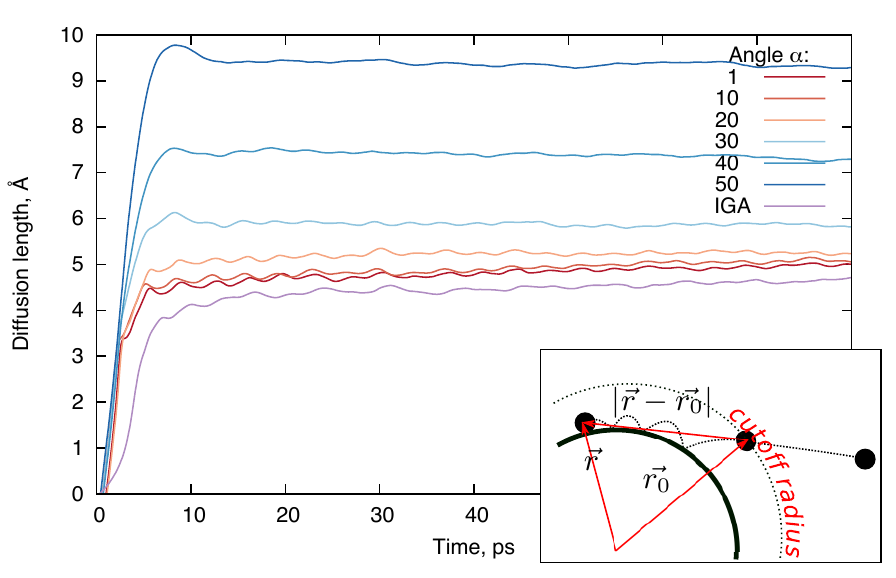}
\end{center}
\caption{Diffusion of incoming particles with different incidence angle on the surface of the cluster. Six curves correspond to high-energy particles with six different angles. The IGA curve corresponds to thermal-energy particles, 0.03~eV. The diffusion length is calculated as a distance between the point where atom reaches the potential  cutoff and the point where the particle is fully thermalized, as shown on the subplot in the right-bottom corner.}
\end{figure}

Fig.~8 compares diffusion of incoming particles on the cluster surface in the IGA and plasma growth processes, points in the figure are obtained by averaging over 160 simulations. One can see that for IGA diffusion does not depend on the incidence angle while plasma growth has a nonlinear dependence for diffusion. In Fig.~8 this nonlinear dependence is fitted with a parabola. \newline

\begin{figure}[h!]
\begin{center}
\includegraphics[width=0.5\textwidth]{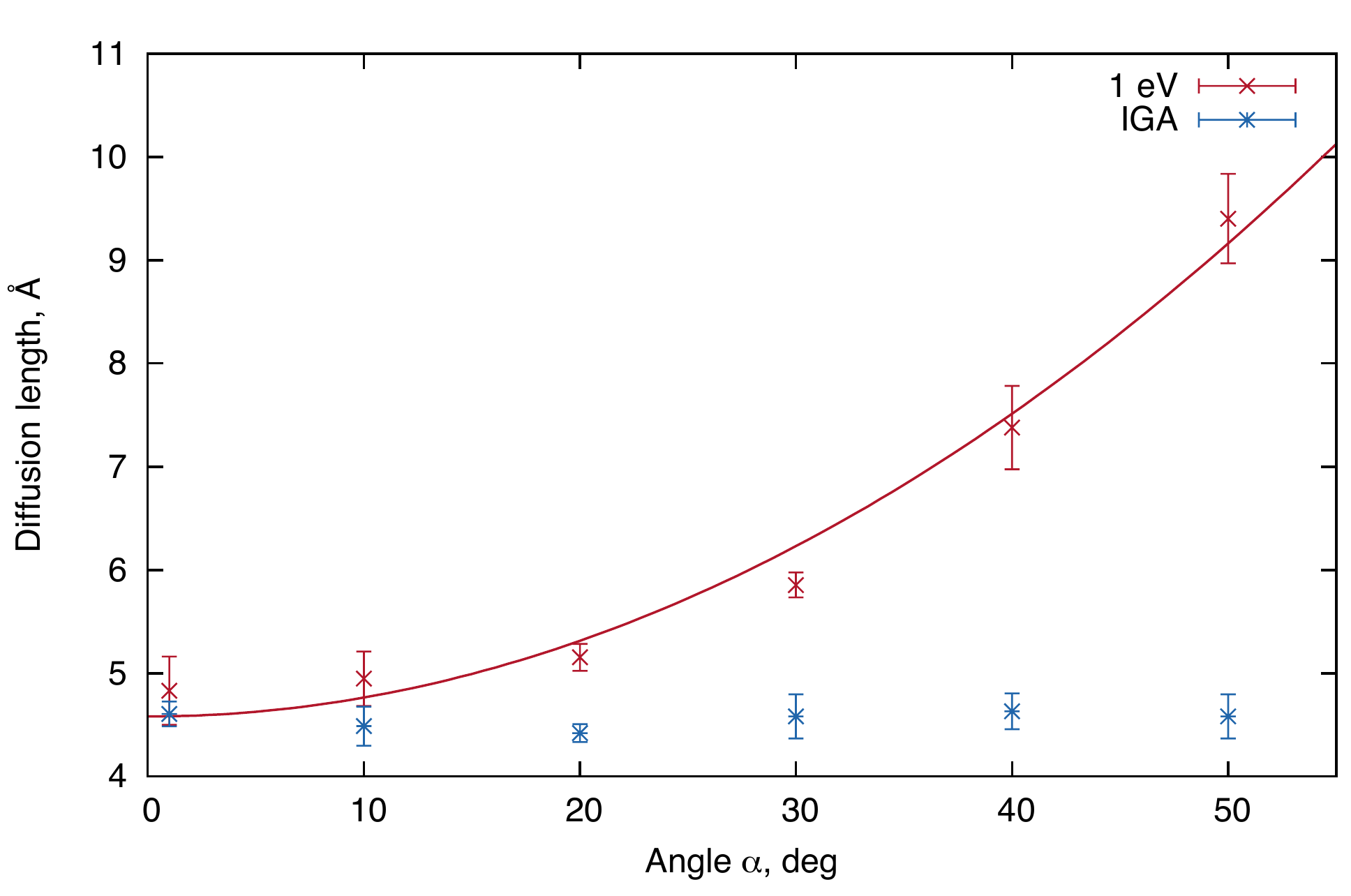}
\end{center}
\caption{Dependence of surface-diffusion length on the incidence angle for incoming particles. The diffusion length of high-energy particles
depends nonlinearly on the incidence angle and is fitted with a parabola while the diffusion length for the IGA process does not show
a dependence.}
\end{figure}

\section{Conclusions}

A model for molecular dynamics simulation of the cluster growth process in a plasma was developed. The model was used to show the difference of the IGA and the growth from ions in a plasma for the example of single events. 
The obtained results clearly show that the angular distribution of incoming particles played an important role and might be taken into account in MD simulations of NP growth with high-energy particles. In particular, we showed how the angular distribution affected the energy transfer from the incoming ion to the NP and surface diffusion. 

Moreover, it was shown that depending on the Coulomb interaction potential there are three possible regimes. The first one corresponds to a very weak Coulomb interaction potential, low or zero charges. In this case incoming particles have a very low velocity near the surface of the cluster and are accelerated towards the cluster by the interatomic interactions and collide with normal incidence. If the Coulomb interaction potential is significant, such as 1 eV, trajectories of the incoming ions bend around the cluster and ions fall on the cluster with angular distribution derived in (6). For the case where the Coulomb interaction potential is high enough, like 10 eV, the incoming ions bombard the cluster with angles close to normal incidence. Our kinetics analysis shows that ions arriving with  normal incidence transfer theirs energy with higher intensity under shorter time. That could lead to a better healing of defects in the NC. On the contrary, particles falling with grazing incidences would not heat the local spot significantly but they have better surface diffusion which can lead to a perfect layer growth at certain conditions.  

Since angular distribution of incoming particles is very sensitive to the initial velocity of particles far from the cluster, we assume that varying the pressure in the growth chamber may significantly change the shape of the angular distribution and thus provide a mean to alter the growth process. For example, CEC ion collection \cite{gatti08} tends to give both lower impact energies and more perpendicular incidences than OML. CEC is definitely promoted by higher pressure. 

\section*{Acknowledgement}
The work was financially supported by the Knut and Alice Wallenberg Foundation through Grant No. 2012.0083. IAA is grateful for the support provided by the Swedish Foundation for Strategic Research (SSF) program SRL Grant No. 10-0026. The supported by the Grant of Russian Federation Ministry for Science and Education (grant No. 14.Y26.31.0005) is gratefully acknowledged. Calculations were performed at the Swedish National Infrastructure for Computing (SNIC) at the National Supercomputer Centre (NSC) in Linköping (Sweden).

\bibliographystyle{plain}

\end{document}